\documentstyle[11pt,IAU207_pasp,twoside,psfig]{article}
\def\edcomment#1{\iffalse\marginpar{\raggedright\sl#1\/}\else\relax\fi}
\reversemarginpar

\begin{document}
\title{Formation of Globular Clusters in Turbulent Molecular Clouds}
\author{Michael P. Geyer \& Andreas Burkert}
\affil{Max-Planck-Institut f\"ur Astronomie, K\"onigstuhl 17, D-69117 Heidelberg, Germany, geyer@mpia-hd.mpg.de}

\begin{abstract}
The formation of massive stellar clusters in turbulent molecular clouds
is investigated. We include artificial star formation and energy feedback of
newly born stars. The obtained systems are not likely to survive. Case 
studies to
determine conditions necessary for forming bound clusters will 
be done in the future.
\end{abstract}

\section{Introduction}

The formation of globular clusters is still an unsolved puzzle.
Massive bound stellar systems can only form by a time coherent collapse of 
large regions of molecular clouds.

Using smoothed particle hydrodynamical simulations (SPH) we investigate 
the formation of stellar clusters in self--gravitating collapsing 
turbulent molecular clouds. 
We mimick star formation by creating N--body particles from gas--particles 
in regions where the the gas density exceeds the Jeans criterion and where 
the gas flow is convergent.

Shortly after the formation of the first stars, the remaining gas 
will be expelled by energy feedback of the massive stars, like
ionizing radiation, stellar winds or supernova explosions. As a result 
of this gas expulsion phase some or all stars in the cluster will also become
unbound. The fraction of finally bound stars is mainly determined by the 
efficiency of star
formation and the timescale of the gas expulsion.

\section{Setup, Star Formation \& Feedback}

We initially start with a homogeneous gas sphere with
a superimposed turbulent velocity field, created as a Gaussian random field
with power spectrum $P(k) \sim k^{-2}$, where $k$ is the wave number.
The turbulence decays and high density regions build up due to
self--gravitation and fragmentation.
The initial parameters are given in table~1. We use
an isothermal equation of state.
The model can be scaled to various initial conditions.

\begin{table}
	\caption{Initial parameters}
	\begin{tabular}{ll}
	\tableline
	gas mass & $M=0.5\, 10^5\, M_\odot$ \\
	radius & $R=50\, \mathrm{pc}$\\
	time unit & $\hat{t}=1.7\,10^7\, \mathrm{yrs}$\\
	temperature &$T=10\, \mathrm{K}$\\
	initial Jeans mass& $M_j=51\, M_\odot$\\
	SPH particle number & $n=26200$\\ \tableline
	\end{tabular}
\end{table}

To follow the formation process of a star cluster from scales as large as 
a molecular cloud down to the stars itself we need a recipe to
mimick star formation. Different approaches concerning 
the inclusion of star formation in SPH
simulations are discussed e.g. by Katz (1992), Bate (1995),
Klessen (2000) and Nakasato, Mori \& Nomoto (2000).
In our simulations, whenever the density of a gas particle exceeds 
the Jeans limit and the flow of the nearest neighbours is convergent, 
a collisionless N--body particle is created with a 
mass according to a 
given intrinsic  star formation efficiency. 
The masses of the gas particle and of its neighbours are reduced accordingly.

After a massive star has formed, it will feed back energy into the interstellar
medium by
ionizing radiation, stellar winds or supernova explosions. A survey of different
methods implementing feedback in SPH is given by Thacker \& Couchman~(2000).
In our simulations each N--body particle contributes energy to its 
surrounding by increasing the thermal energy of the neighbouring 
gas particles at a constant rate.

\section{Dynamical Evolution}

\begin{figure}
	\psfig{figure=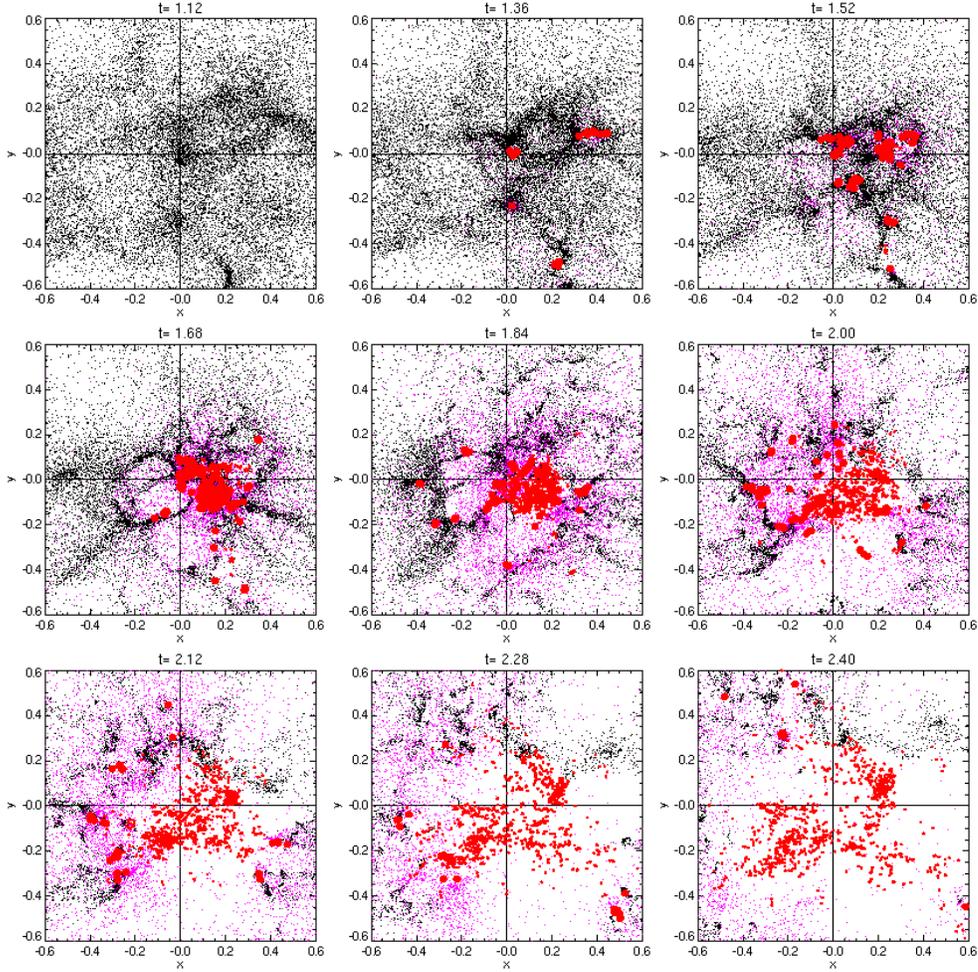,width=13.0cm}
	\caption{Time evolution projected onto the $x-y$-plane. 
	Black and pink dots: cold and hot gas particles.
	Red dots: N--body particles.}
\end{figure}

Fig.~1 shows a time sequence of one simulation.
The turbulence decays and the 
cloud starts collapsing. At $t=1.36$ the first
stars form in high density regions. 
Their energy feedback expells the surrounding gas and
causes star formation in the compressed medium in front of 
outwards propagating gas shells.
The newly born stars gain
high velocities and are not bound in the system of stars and gas.

\begin{figure}
	\centerline{\hbox{
	\psfig{figure=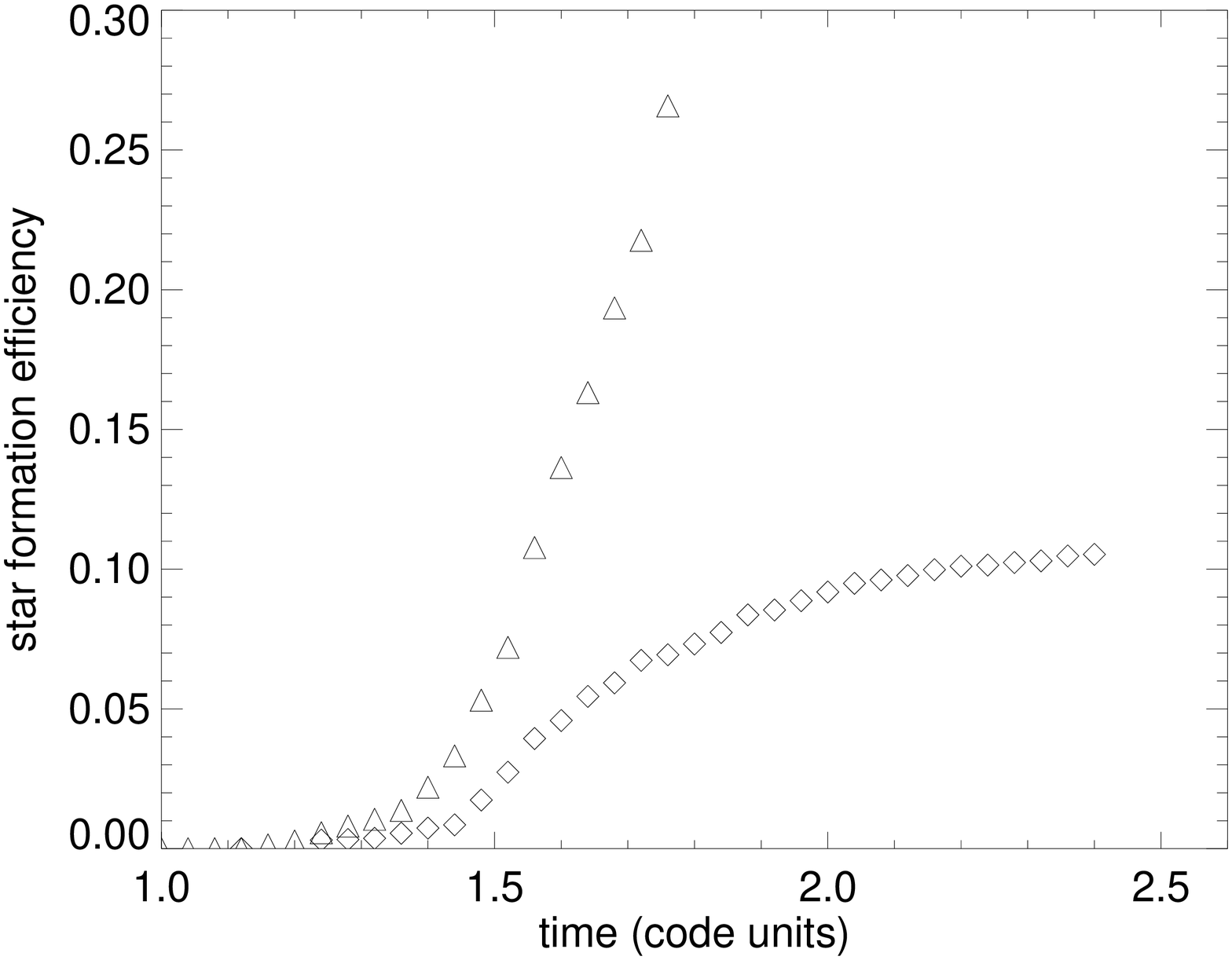,width=5.625cm,height=4.5cm}}}
	\caption{Time evolution of the star formation efficiency 
	(ratio of the mass in stars to initial gas cloud mass) without (triangles)
	and with (diamonds) energy feedback.}
\end{figure}

Fig.~2 compares the star formation efficiency with and without energy feedback.
In accordance with observations, our model including
energy feedback shows only low star formation
efficiencies. Thus, the star cluster
that has formed is not likely to survive (Lada, Margulis \& Dearborn 1984,
Geyer \& Burkert 2001).
This result is independent of the detailed assumption of feedback or the initial
condition. It demonstrates that the formation of massive bound
clusters requires conditions that are not typically found in molecular clouds.
In future simulations we will investigate the requirements necessary for
forming massive bound clusters.

\end{document}